\documentclass{aa} 
\usepackage{graphicx}
\usepackage{float}
\usepackage{txfonts}
\usepackage{multirow}
\usepackage{natbib}
\bibpunct{(}{)}{;}{a}{}{,}
\usepackage{hyperref}                         
\hypersetup{
    colorlinks=true,
    linkcolor=blue,
    citecolor=blue
    }

\newcommand{\hrieuv}{HRI$_{\textrm{EUV}}$}
\newcommand{\hrilya}{HRI$_{\textrm{Lya}}$}

\def\arcsec{$^{\prime\prime}$}

\usepackage{xcolor}

\begin{document}

\title{Imaging and spectroscopic observations of extreme-ultraviolet brightenings using EUI and SPICE on board Solar Orbiter}

\titlerunning{}

\author{
        Ziwen Huang\inst{\ref{i:mps}}\fnmsep\thanks{Corresponding author: Ziwen Huang \email{huangz@mps.mpg.de}}
        \and
         L. Teriaca\inst{\ref{i:mps}}
        \and
        R. Aznar Cuadrado\inst{\ref{i:mps}}
        \and
        L. P. Chitta\inst{\ref{i:mps}}
        \and
        S. Mandal\inst{\ref{i:mps}}
        \and
        H. Peter\inst{\ref{i:mps}}
        \and
        U. Sch\"{u}hle\inst{\ref{i:mps}}
        \and
        S.K. Solanki\inst{\ref{i:mps}}
        \and
        F. Auch\`{e}re\inst{\ref{i:ias}}
        \and
        D. Berghmans\inst{\ref{i:rob}}
        \and
        É. Buchlin\inst{\ref{i:ias}}
        \and
         M. Carlsson\inst{\ref{i:rocs},\ref{i:uio}}
          \and
         A. Fludra\inst{\ref{i:ral}}
         \and        
         T. Fredvik\inst{\ref{i:rocs}, \ref{i:uio}}
        \and
        A. Giunta\inst{\ref{i:ral}}
        \and
        T. Grundy\inst{\ref{i:ral}}
        \and
         D. Hassler\inst{\ref{i:swri}}
        \and
        S. Parenti\inst{\ref{i:ias}}
        \and
        F. Plaschke\inst{\ref{i:tub}}
}

\color{black}

  \institute{
            Max Planck Institute for Solar System Research, Justus-von-Liebig-Weg 3, 37077 G\"ottingen, Germany\label{i:mps}
         \and
            Université Paris-Saclay, CNRS, Institut d'Astrophysique Spatiale, 91405, Orsay, France\label{i:ias}
           \and 
            Solar-Terrestrial Centre of Excellence -- SIDC, Royal Observatory of Belgium, Ringlaan -3- Av. Circulaire, 1180 Brussels, Belgium\label{i:rob}
           \and 
            Rosseland Centre for Solar Physics, University of Oslo,  P.O. Box 1029 Blindern, NO-0315 Oslo, Norway\label{i:rocs}
           \and 
            Institute of Theoretical Astrophysics, University of Oslo,  P.O. Box 1029 Blindern, NO-0315 Oslo, Norway\label{i:uio}
            \and      
            RAL Space, UKRI STFC, Rutherford Appleton Laboratory, Didcot, UK\label{i:ral}
            \and
            Southwest Research Institute, Boulder, CO, USA\label{i:swri}
            \and
            Institut f\"{u}r Geophysik und extraterrestrische Physik, Technische Universit\"{a}t Braunschweig,
            Mendelssohnstrasse 3, 38106 Braunschweig, Germany\label{i:tub}
}

 \date{Received XX, YY, 2023; accepted XX, YY 2023}

\abstract
{The smallest extreme-ultraviolet (EUV) brightening events that were detected so far, called campfires, have recently been uncovered by the High Resolution EUV telescope (\hrieuv), which is part of the Extreme Ultraviolet Imager (EUI) on board Solar Orbiter. \hrieuv\ has a broad bandpass centered at 17.4~nm that is dominated by Fe~{\sc ix} and Fe~{\sc x} emission at about 1~MK.}
{We study the thermal properties of EUI brightening events by simultaneously observing their responses at different wavelengths using spectral data from the Spectral Imaging of the Coronal Environment (SPICE) also on board Solar Orbiter and imaging data from EUI.}
{We studied three EUI brightenings that were identified in \hrieuv\ data that lie within the small areas covered by the slit of the SPICE EUV spectrometer. We obtained the line intensities of the spectral profiles by Gaussian fitting. These diagnostics were used to study the evolution of the EUI brightenings over time at the different line-formation temperatures.}
{We find that (i) the detection of these EUI brightenings is at the limit of the SPICE capabilities. They could not have been independently identified in the data without the aid of \hrieuv\ observations. (ii) Two of these EUI brightenings with longer lifetimes are observed up to Ne~{\sc viii} temperatures (0.6~MK). (iii) All of the events are detectable in O~{\sc vi} (0.3~MK), and the two longer-lived events are also detected in other transition region (TR) lines. (iv) In one case, we observe two peaks in the intensity light curve of the TR lines that are separated by 2.7~min for C~{\sc iii} and 1.2~min for O~{\sc vi}. The Ne~{\sc viii} intensity shows a single peak between the two peak times of the TR line intensity.}
{Spectral data from SPICE allow us to follow the thermal properties of EUI brightenings. Our results indicate that at least some EUI brightenings barely reach coronal temperatures.}

\titlerunning{EUI and SPICE observations of EUI brightenings}
\authorrunning{Z. Huang et al.}

\maketitle

\section{Introduction}
\label{sec-intro}

The smallest coronal transient extreme-ultraviolet (EUV) brightening events that were observed so far, so-called campfires, were discovered by 
\citet{2021A&A...656L...4B} % Berghmans et al. 2021 
using data from the High Resolution Extreme Ultraviolet telescope (\hrieuv), which is part of the Extreme Ultraviolet Imager 
\citep[EUI;][]{2020A&A...642A...8R} % Rochus et al. EUI instr. paper.
on board the ESA and NASA Solar Orbiter mission
\citep{2020A&A...642A...1M}. % Solar Orbiter paper.
In EUI data taken on 30 May 2020 during the commissioning phase of Solar Orbiter, more than one thousand campfires were identified in the about 4-minute observation, and their properties were reported. Although the statistical analysis is based on a limited data sample, it shows that the campfires usually display one of two morphologies, dot-like and loop-like. Most of them are small-scale and short-lived events. Even loop-like campfires are not longer than 4~Mm in their elongated directions. Most events last some tens of seconds, and the longest-lived events approach the limit set by the duration of the observations, about 200~s. %{\color{blue}\citet{Nelson_tbs_CF} % Nelson et al. cf submitted to this issue
%systematically studied EUI data taken on 8 March for about 29.4~min and got similar results. However, since they have longer observation duration, they have also found brightenings with longer life time (larger than 20~min). A correlation between sizes, intensities and lifetimes is also reported. }
The high spatial and temporal resolution of \hrieuv\ enables us to study the small EUV brightenings that are barely detected in other observations, for example, from Atmospheric Imaging Assembly \citep[AIA;][]{2012SoPh..275...17L} % AIA paper Lemen et al. 2012
on the Solar Dynamics Observatory \citep[SDO;][]{2012SoPh..275....3P}. % Pesnell et al. 2012
We use the name "EUI brightenings" in this paper to refer to these small-scale EUV brightenings seen by EUI.

\begin{figure*}[tbp]
\centering 
\includegraphics[width=1.0\textwidth]{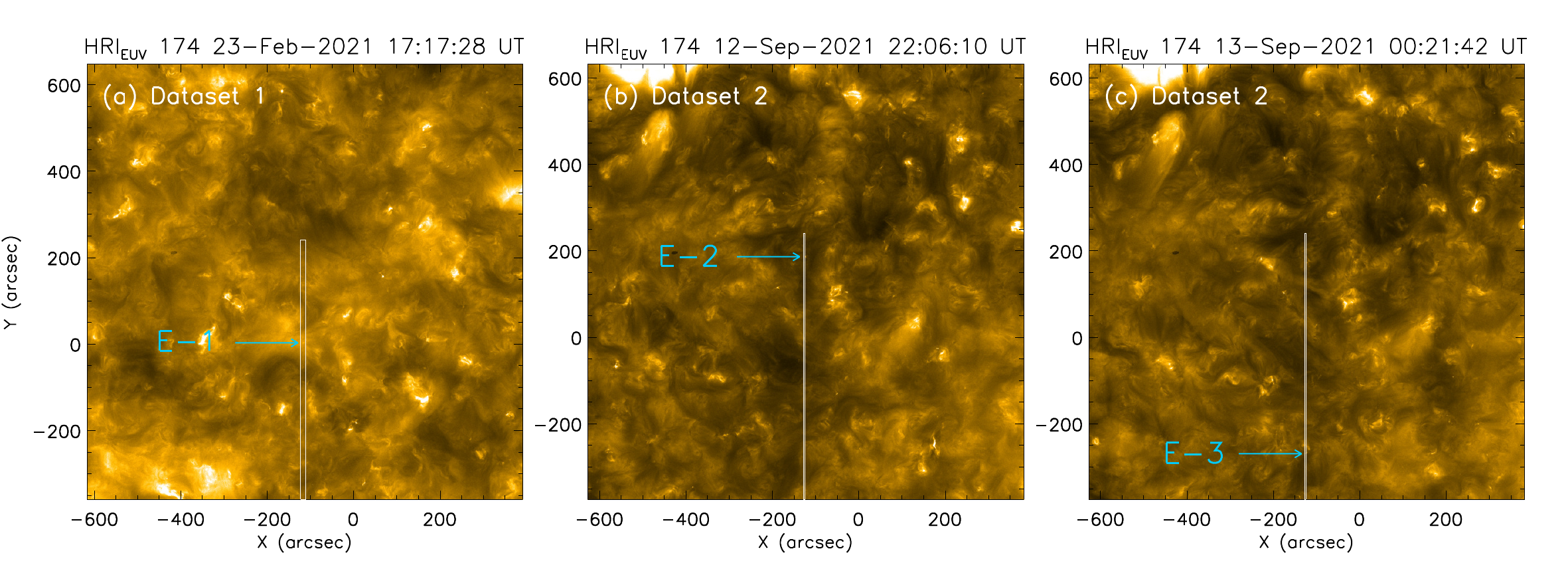}
\caption{\hrieuv\ images of the solar corona taken from the analyzed data sets. (a) Image obtained on 23 February 2021 at 17:17:28 UT (data set 1). The region that is also covered by SPICE slit (3-step) is marked by the white box. (b) Image obtained on 12 September 2021 at 22:06:10 UT (data set 2). The region also covered by the SPICE slit (sit and stare) is marked by the white box. (c) Image obtained on 13 September 2021 at 00:21:42 UT (data set 2).  The region also covered by the SPICE slit (sit and stare) is marked by the white box. Blue arrows in all panels point to the positions of EUI brightenings.}
\label{FigC}
\end{figure*}

The discovery of EUI brightenings extends the boundary of EUV brightening event sizes to a smaller scale. Brightening events are transient structures that are normally observed in UV and EUV bandpasses in active regions \citep[e.g.,][]{1984ApJ...283..879P,1992PASJ...44L.147S,1999SoPh..186..207B}, % Porter, Toomre, and Gebbie 1984, Shimizu et al. 1992, Berghmans 1999
in the quiet Sun, and in coronal holes \citep[e.g.,][]{1990ApJ...352..333H,1991ApJ...382..667H,1998A&A...336.1039B,2003A&A...398..775M}. % Habbal 1990, 1991, Berghman 1998, Madjarska 2003
A systematic study conducted by \citet{2022A&A...661A.149P} showed that their distribution does not vary throughout the solar cycle.
EUV brightenings have been studied in past decades in an attempt to identify their potential role in coronal heating. As proposed by \citet{1974ApJ...190..457L}, %Levine 1974
the total energy released by small reconnection episodes could make an important contribution to the heating of the solar atmosphere. Based on more flares that were observed at smaller scales,
\citet{1981ApJ...244..644P, 1983ApJ...264..642P, 1988ApJ...330..474P} %Parker 1981,1983,1988
suggested that a multitude of small flares occurring from magnetic reconnection could provide the energy to keep the corona at a high temperature. He also introduced the term nanoflare, which refers to energy-release events of about 10$^{17}$\,J. This term was later also used for small impulsive brightening events in observations. The relation between the occurrence of solar flares and their total released energy can be described with a power-law distribution. In past decades, many studies were conducted to determine the index of these power laws \citep[e.g.,][]{1991SoPh..133..357H,Benz_2002,Aschwanden_2002}. %Hudson 1991, Benz & Krucker 2002, Aschwanden & Parnell 2002  
\citet{1991SoPh..133..357H} pointed out that the energy provided by small events can dominate the total heating energy only when the slope of the power law is steeper than 2. 
However, a flatter slope has been concluded based on EUV and X-ray observations \citep[e.g.,][]{2000ApJ...535.1047A,2013ApJ...776..132A}. %Aschwanden et al. 2000, Aschwanden & Shimizu 2013
\citet{2016A&A...591A.148J} % Joulin et al. 2016
discussed that even with a power-law index higher than $-2$, brightening events can only provide a small contribution if the energy range is too small or the occurrence frequency is too low.

In nanoflare heating models, the heating mechanism is closely related to the reconnection of magnetic field lines \citep[e.g.,][]{2006SoPh..234...41K,2015RSPTA.37340256K}. % Klimcluk 2006,2015 
The power-law distributions in EUI brightening in projected area, duration, and total intensity may support the idea that they are smaller-scale variants of solar flares, that is, micro- or nanoflares. 
Observations with the \hrilya\ telescope of EUI, which is dominated by emission from the Lyman-$\alpha$ line of hydrogen, show that most EUI brightenings are located at the boundaries of the chromospheric network, indicating that their origin might be related to magnetic fields. This would also be  consistent with their prevalent loop-like form \citep{2021A&A...656L...4B}. % Berghmans et al. 2021
The relation between EUI brightenings and magnetic reconnection, most likely driven by motions at the footpoints of small loops, can also be studied by investigating the magnetic features underlying them using the data obtained with the Polarimetric and Helioseismic Imager on board Solar Orbiter \citep[SO/PHI;][]{2020A&A...642A..11S}, for example% PHI paper Solanki et al. 2020 
. \citet{2022A&A...660A.143K} % F. Kahil et al. 2022
showed that most EUI brightenings are indeed associated with episodes of magnetic flux cancellation at the footpoints of small loops. On the other hand, they also reported that about 25\% of the EUI brightenings they studied are not associated with evident magnetic activity in the photosphere, which may imply that other heating mechanisms also play a role. The connection of EUI brightenings with opposite-polarity magnetic features has also been found \citep{2021ApJ...921L..20P} % Panesar et al. 2021) 
using data from the Helioseismic and Magnetic Imager \citep[HMI;][]{2012SoPh..275..207S}. % Scherrer et al. 2012).
In magnetohydrodynamics (MHD) simulations conducted by \citet{2021A&A...656L...7C}, % Yajie Chen et al. 2021 
most brightening events with characteristics comparable to the observed EUI brightenings appear to be driven by magnetic reconnection. The correlation between the occurrence of EUI brightenings and magnetic field activities may reveal that these EUI brightenings share similar formation and evolution mechanisms with larger brightenings or flares.

Temperature is another important property to determine how small structures contribute to coronal heating. Differential emission measure (DEM) analysis helps to estimate temperatures of events \citep{2012A&A...539A.146H, 2015ApJ...807..143C}. %Hannah & Kontar 2012, Cheung M.C.M 2015 
As shown in results from a DEM analysis with data from AIA
on board the SDO, the distribution of DEM-weighted temperatures in EUI brightenings and similar small EUV brightenings peaks at $\log T\rm{[/K]}\sim6.12$, which is the typical coronal temperature \citep{2021A&A...647A.159C,2021A&A...656L...4B}. % Chitta et al. 2021 and Berghmans et al. 2021. 

Even though the DEM method is effective and widely used to diagnose the temperature information of coronal structures of different sizes, it contains uncertainty from the assumptions about the spectral response of the passbands of broadband imagers such as AIA and \hrieuv. The DEM inversion of AIA data can be systematically biased toward 1~MK for multithermal plasma \citep{2012ApJS..203...26G}, % Guennou et al. 2012 
which may explain why 1~MK is found in EUI brightenings.

Thus, spectral data are necessary to study the temperature evolution of solar events. Studies have been conducted of small solar events such as bright points, small jets, or explosive events by using lines with different formation temperatures (in the transition region (TR) and in the corona) recorded by spectrometers. For instance, \citet{1997Natur.386..811I} %Innes et al. 1997
reported bidirectional jets resulting from magnetic reconnection using data from the Solar Ultraviolet Measurements of Emitted Radiation \citep[SUMER;][]{1995SoPh..162..189W} %SUMER paper
spectrometer on board the Solar and Heliospheric Observatory (SOHO). Non-Gaussian line profiles of O~{\sc vi} (103.193~nm) show mass flows in small structures in the TR \citep{2004A&A...427.1065T}. %L. Teriaca 2004 
\citet{2008ApJ...681L.121T} % H. Tian 2008
studied both cool and hot components in an EUV bright point with ten lines. All these studies enriched our understanding of the complicated TR and corona. However, the thermal properties of EUI brightenings have not been explored with spectroscopic data so far. 

\begin{figure*}[tbp]
\centering 
\includegraphics[width=0.8\textwidth]{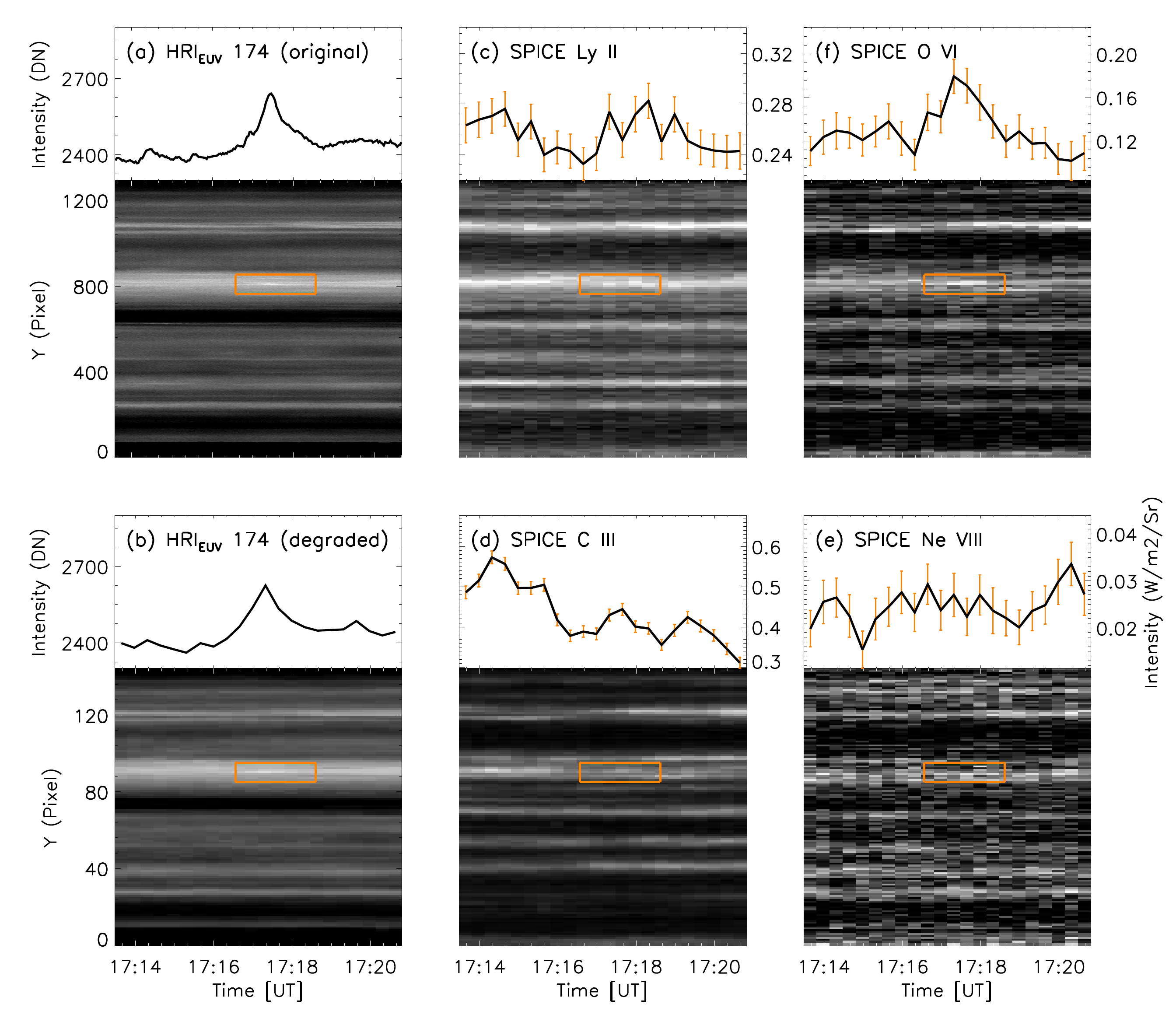}
\caption{Multithermal view of an EUI brightening (E-1) obtained on 23 February 2021. Panels (a) and (b): Time-slice plots of \hrieuv\ images with full and degraded (to SPICE spatial and temporal) resolution. Panels (c)-(f): Time-slice plots of the intensity from the Gaussian fitting results of Ne~{\sc viii}, C~{\sc iii}, Ly$\beta,$ and O~{\sc vi}. Orange boxes mark the position in which the EUI brightening appeared. The curves at the top of each panel show the intensities at this position vs. time. The error bars were calculated from the errors of the fitting parameters and are plotted in orange.
}

\label{FigE11}
\end{figure*}

In this study, we first investigate EUI brightenings using both imaging and spectral data. The instruments \hrieuv\ and SPectral Imaging of the Coronal Environment 
\citep[SPICE;][]{2020A&A...642A..14S, 2021A&A...656A..38F} %Spice Instr. paper 
simultaneously observed the same region on the Sun in 2021 during the cruise phase of Solar Orbiter. This study investigates the temperature properties of EUI brightenings by combining EUV images from \hrieuv\ and spectra from SPICE. Three EUI brightenings from two periods of time are studied in this work. An overview of the data set we used is shown in Section \ref{sec-obs}. In Section \ref{sec-ana} we explain the methods we used for analysis. Detailed descriptions of all three cases and results are presented in Section \ref{sec-res}, followed by discussions and our conclusions in Section \ref{sec-con}.

\begin{figure*}[tbp]
\centering 
\includegraphics[width=1.0\textwidth]{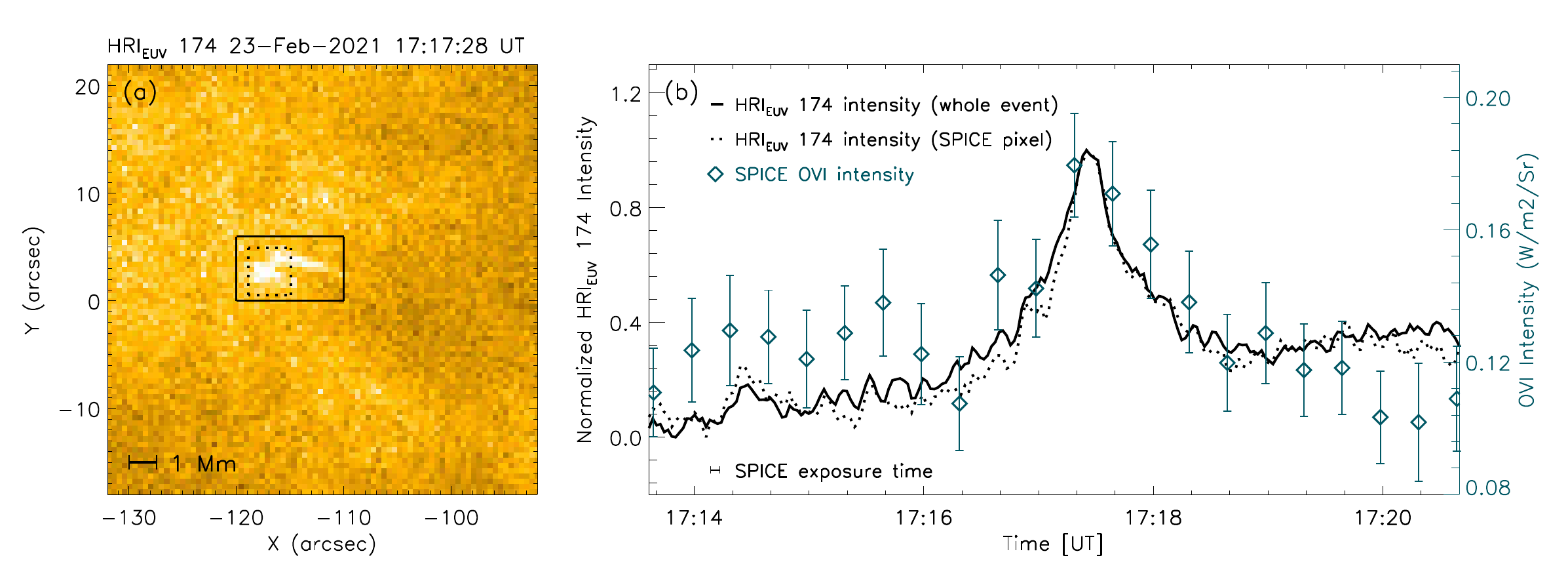}
\caption{EUI brightening (E-1). Panel (a): \hrieuv\ image taken on 23 February 2021 at 17:17:28~UT, at about the time when the intensity of the EUI brightening peaked. The smaller dashed box shows the region covered by one SPICE-binned pixel ($4\text{\arcsec}\times 4.392$\arcsec) when the slit was at the second spatial step of the short raster. The O~{\sc vi} intensity at this pixel (blue diamonds) in panel (b) is calculated from Gaussian fits to the line in the SPICE data. The error bars were calculated from the uncertainties of the fitting parameters. This region was also used to calculate the \hrieuv\ intensity (dashed line) in panel (b). The larger solid box shows the region we used to compute the light curve (solid line) shown in panel (b). The horizontal bar in the lower left corner shows the SPICE exposure duration, which is about 4.7~s. An animation of panel (a) is available.}
\label{FigE12}
\end{figure*}

\begin{figure}[tbp]
\centering 
\includegraphics[width=1.\hsize]{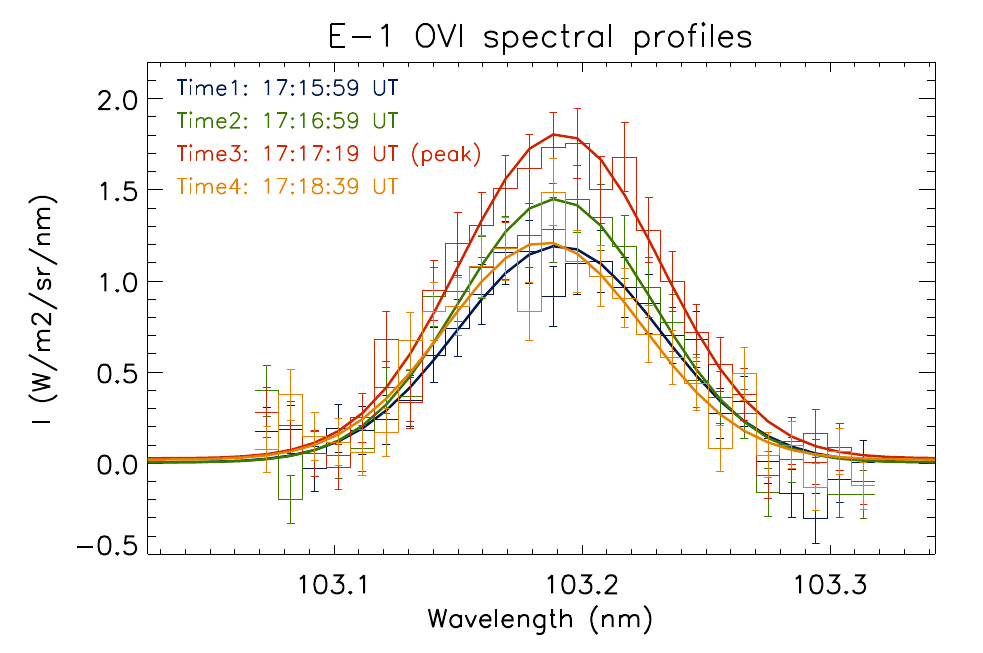} 
\caption{EUI brightening (E-1). O~{\sc vi} spectral profiles (histograms with error bars) of one binned SPICE pixel marked by the dashed white box in panel (a) of Fig. \ref{FigE12} and the Gaussian fitting results (thick lines) at four different times during the lifetime of E-1.}
\label{FigE13}
\end{figure}

\section{Observations}
\label{sec-obs}
For the purpose of studying their temperature properties, we selected EUV brightening events that can be identified in \hrieuv\ data and are also covered by the slit of SPICE. In this study, we focus on three cases from two different parts of the spacecraft orbit. Information about the data sets is included in Table~\ref{tab1}, where all times are the starting times of the observations. The EUI data we used in this study are available in the EUI data release 5.0\footnote{\url{https://doi.org/10.24414/2qfw-tr95}} , and the SPICE data are included in the SPICE data release 3.0\footnote{\url{https://doi.org/10.48326/idoc.medoc.spice.3.0}}. Figure~\ref{FigC} shows overviews of the \hrieuv\ field of view that is covered in these two data sets. 

On 23 February 2021, when Solar Orbiter had a heliocentric distance of 0.53~AU and the angular separation between Earth and Solar Orbiter was about 140 degrees, \hrieuv\ first provided a sequence with a cadence of 2~s at 17.4~nm from 17:13:25~UT to 17:20:59~UT. \hrieuv\ images have a high angular image scale of 0.492\arcsec/pixel. At the heliocentric distance of 0.53~AU, this corresponds to about 190~km on the Sun (at a resolution of about 380~km). At about this time, SPICE took both context and small high-cadence rasters. The two context rasters are obtained by stepping the 4\arcsec\,wide (1500~km on the Sun) slit in 192 steps of 4\arcsec\,with an exposure time of 20~s at each position, starting at 15:44:00~UT (opening context raster) and at 17:33:43~UT (closing context raster). From 16:50:31~UT to 17:23:31~UT, we have a small raster from SPICE every 20~s. For each small raster, the 4\arcsec\,wide slit scanned over three spatial positions in steps of 4\arcsec\,and with 4.7~s exposures. The pixel scale of SPICE along the slit is 1.098\arcsec\,(420~km).

The other data sets were obtained on 12 and 13 September 2021, when Solar Orbiter was 0.59~AU away from the Sun and the angular separation between Earth and Solar Orbiter was about 50 degrees. 
This time, SPICE observed in sit-and-stare mode with an exposure time of 10~s, which means that the slit was kept at a fixed position on the Sun. The pixel scale of SPICE in this data set is about 470~km. Three sets of image sequences were taken by \hrieuv, but only the first and third sets have a high time cadence of 2~s. The cadence of the second sequence was set to 1~min, which is not ideal for detecting an EUI brightening, considering the short life span of the majority of these events. All \hrieuv\ images were taken with an exposure time of 1.65~s and a pixel scale of 0.492\arcsec\,(210~km). In this study, we used level 2 data, which are corrected for instrumental effects. The \hrieuv\ level 2 data are returned in units of DN (digital numbers) per second, and SPICE level 2 data are in units of $\text W/\text m^2/\text{sr}/\text{nm}$. 

\begin{table*}[ht]
    \centering
    \caption{Overview of the observations.}
    \label{tab1}
    \resizebox{\textwidth}{!}{
    \begin{tabular}{*{9}{c}}
      \hline
      \multirow{2}*{Date} & \multicolumn{3}{c}{\hrieuv} & \multicolumn{4}{c}{SPICE} \\
      \cline{2-4}\cline{5-9}
      & Time & Cadence & Exposure & Time (UT) & Cadence & Exposure & Type & Version\\
      \hline
       \multirow{3}*{23-Feb-2021} & \multirow{3}*{17:13:25-17:20:59} & \multirow{3}*{2~s} & \multirow{3}*{1.65~s} & 15:44:00 & \multirow{2}*{N/A} & \multirow{2}*{20~s} & \multirow{2}*{Context rasters} & \multirow{2}*{V10}\\
       \cline{5-5}
       &  &  &  & 17:33:43 &  &  & \\
       \cline{5-9}
       &  &  &  & 16:50:31-17:23:31 & 20~s & 4.7~s & Small rasters & V10\\
       \hline
       \multirow{3}*{12/13-Sep-2021} & 22:00:59-22:14:59 & 2~s & \multirow{3}*{1.65~s} & 22:04:19-23:12:09 & \multirow{3}*{10.2~s} & \multirow{3}*{10~s} & \multirow{3}*{Sit and stare} & V08\\
       \cline{2-3}\cline{5-5}\cline{9-9}
       & 23:05:00-23:19:00 & 1~m &  & 23:12:20-00:20:10 &  &  &  & V12\\
       \cline{2-3}\cline{5-5}\cline{9-9}
       & 00:11:51-00:24:59 & 2~s &  & 00:20:20-01:28:10 &  &  &  & V05\\
      \hline
    \end{tabular}}
\end{table*}

Since the SPICE slit covers just a narrow region, only three EUI brightenings are identified in the EUI and SPICE data, which we call E-1, E-2, and E-3 hereafter for convenience. These EUI brightenings are all very small, with an area smaller than $5\,\text{Mm}^2$, which is at the limit of the detection capabilities of SPICE. On the temporal scale, E-1 existed for about 40~s, and the other two cases were observed over several minutes. More detailed timing information can be found in Table \ref{tab2}.

\begin{table*}[ht]
    \centering
    \caption{Event characteristics}
    \label{tab2}
    \begin{tabular}{*{6}{c}}
      \hline
      Event & Start time (UT) & Peak time (UT) & Duration & Area ($Mm^2$) & SPICE spectral lines \\
      \hline
      E-1 & 17:17:04 & 17:17:28 & 50~s & 2.2 & Mg~{\sc ix}, Ne~{\sc viii}, C~{\sc iii}, Ly$\beta$, O~{\sc vi}\\
      E-2 & 22:04:24 & 22:06:10 & 6~m & 3.8 & N~{\sc iv}, Ne~{\sc viii}, S~{\sc v}, O~{\sc iv} (78.7, 79~nm), C~{\sc iii}, Ly$\beta$, O~{\sc vi}\\
      E-3 & 00:20:16 & 00:21:40 & 4~m & 4.5 & N~{\sc iv}, Ne~{\sc viii}, S~{\sc v}, O~{\sc iv} (78.7, 79~nm), C~{\sc iii}, Ly$\beta$, O~{\sc vi}\\
      \hline
    \end{tabular}
\end{table*}

\section{Data analysis}
\label{sec-ana}

Table \ref{tab2} shows  that E-1 is covered by Mg~{\sc ix}~70.602~nm ($\log T\rm{[/K]}=6.0$) and Ne~{\sc viii}~77.042~nm ($\log T\rm{[/K]}=5.8$) and by TR lines such as C~{\sc iii}~97.703~nm ($\log T\rm{[/K]}=4.8$), Ly$\beta$~102.572~nm ($\log T\rm{[/K]}=4.0$), and O~{\sc vi}~103.193~nm ($\log T\rm{[/K]}=5.5$). Mg~{\sc ix} is not used in this case because the signal-to-noise ratio is very low. For the other two cases, the Short Wavelength (SW) detector of SPICE also provides N~{\sc iv}~76.515~nm ($\log T\rm{[/K]}=5.1$), S~{\sc v}~78.647~nm ($\log T\rm{[/K]}=5.2$), O~{\sc iv}~78.772~nm ($\log T\rm{[/K]}=5.2$), and O~{\sc iv}~79.019~nm ($\log T\rm{[/K]}=5.2$) in addition to the Ne~{\sc viii} line. SPICE spectra are dominated by the instrumental profile, which can be reasonably fitted by a Gaussian function. We used this to retrieve the integrated line radiance. Before the Gaussian fit process, the original data were prepared to remove the effect of pixels with high negative values mostly due to cosmic-ray spikes that affect the single dark frame that is used for the correction and bright spikes that are due to cosmic rays hitting the detector during observations. All lines can be fitted with a single Gaussian function, except for the spectral feature formed by the S~{\sc v}~78.647~nm, O~{\sc iv}~78.772~nm, and O~{\sc iv}~79.019~nm lines. At the SPICE spectral resolution, these lines appear to be substantially blended. This feature was consequently fitted with a triple Gaussian function. The O~{\sc iv}~79.019~nm line was not used for further analysis in our cases, because it is very close to the edge of the detector and is thus not reliable. Every O~{\sc iv} line mentioned below refers to the O~{\sc iv}~78.772~nm line.

In the fitting process, we used custom software based on the the \texttt{xcfit\_block.pro} routine from Solarsoft, taking the prepared data and the corresponding estimated error (see Appendix~\ref{error}) as inputs to obtain the parameters of the Gaussian function(s) (amplitude, line center, and position) and of the background at each point. 
We considered the background to be a constant along the wavelength in the window of each line. To improve the signal-to-noise ratio of the data, we binned over 4 spatial pixels along the direction of the slit. Since the width of the slit is 4\arcsec, about four times the imaging spatial scale, this binning led to square pixels in the resulting raster scans. Moreover, the spatial resolution of SPICE is estimated to be 
about 6 to 7\arcsec\,, which means that we do not lose substantial information on the target EUI brightenings because of this binning. In the following discussion, all references to SPICE data assume the 4-pixel binning along the slit direction for an effective plate scale of $4\text{\arcsec} \times 4.392$\arcsec (hereafter one SPICE-binned pixel), which is about $1540 \times 1690$~km for event E-1 and $1710 \times 1880$~km for events E-2 and E-3. From the fitting results, we obtained the radiance at every pixel after binning, as described above. 

Unlike large-scale and long-lived structures, which can clearly be identified in \hrieuv\ images and spectral scanning images, EUI brightenings are difficult to distinguish in SPICE scanning data without the aid of \hrieuv\ data. 
In order to study the EUI brightening signatures in SPICE data, it is therefore first necessary to accurately align the the data from the two instruments. The detailed alignment method can be found in Appendix~\ref{data-align}, where we apply a two-step approach and make use of time-series data.

By carefully aligning \hrieuv\ and SPICE data, we determined the positions of the EUI brightenings, as identified in \hrieuv\ data, in the SPICE data. An easy way of comparing the signature of EUI brightenings in \hrieuv\ and SPICE data is to reduce the resolution of \hrieuv\ to that of SPICE. However, to confirm whether the SPICE slit scanned the EUI brightening structure during the simultaneous observation time period, we used the \hrieuv\ data at their full resolution because this shows the evolutionary trace of EUI brightenings in time-slice images more clearly. Since data set 1 consists of rasters with three spatial steps, we are able to obtain a time-slice plot at each spatial position. In data set 2, we obtain the time-slice plot at one single spatial position, but with higher temporal resolution.

\begin{figure*}[tbp]
\centering 
\includegraphics[width=1.0\textwidth]{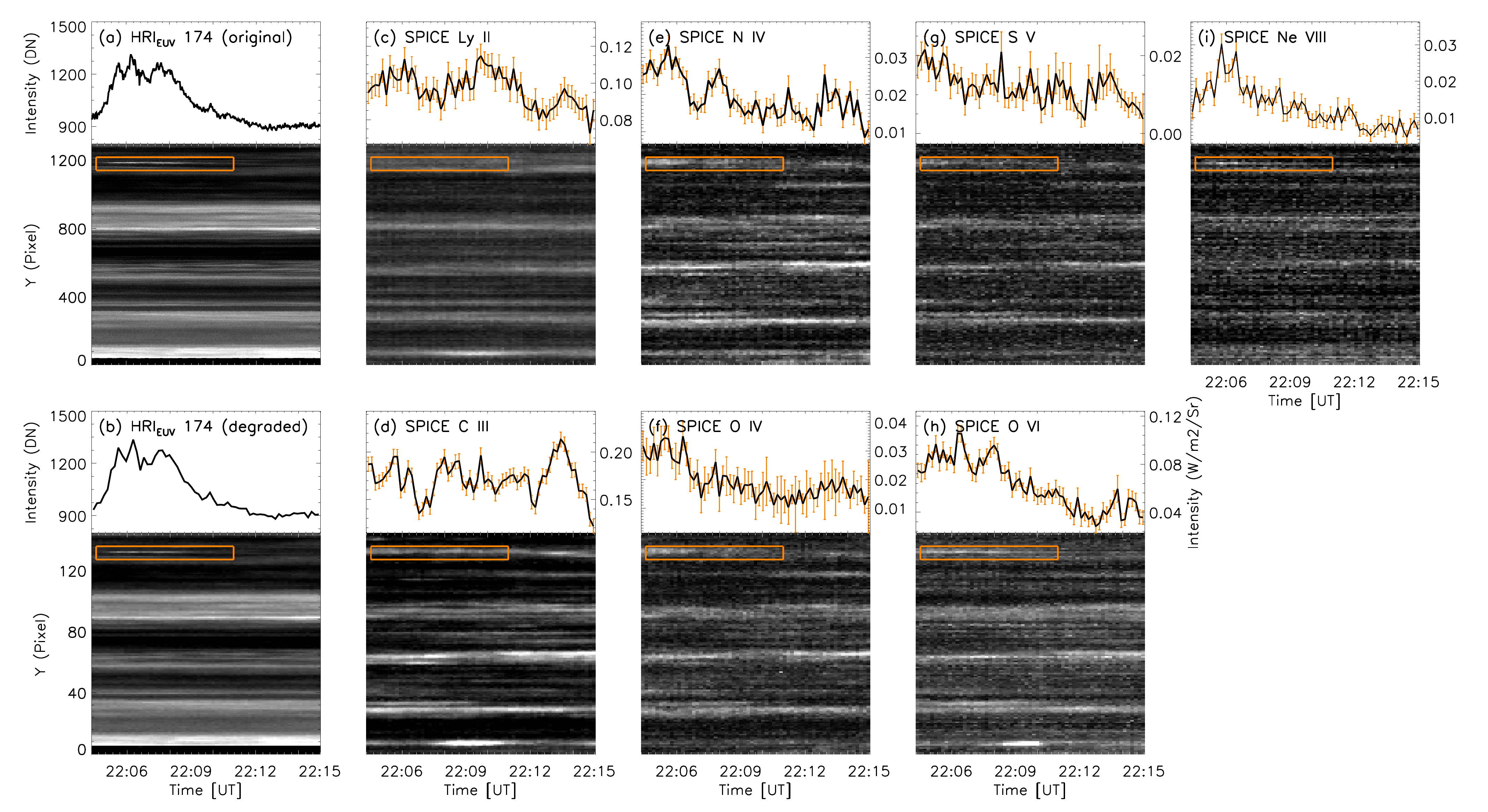}
\caption{Multithermal view of an EUI brightening (E-2). Same as Fig.~\ref{FigE11}, but for E-2 on 12 September 2021. It shows time-slice plots and light curves from the \hrieuv\ data at full and reduced (to SPICE) resolution and from Gaussian fitting results of the Ly$\beta$, C~{\sc iii}, N~{\sc iv}, O~{\sc iv}, S~{\sc v}, O~{\sc vi,} and Ne~{\sc viii} spectral lines.
}
\label{FigE21}
\end{figure*}

\begin{figure*}[tbp]
\centering 
\includegraphics[width=1.0\textwidth]{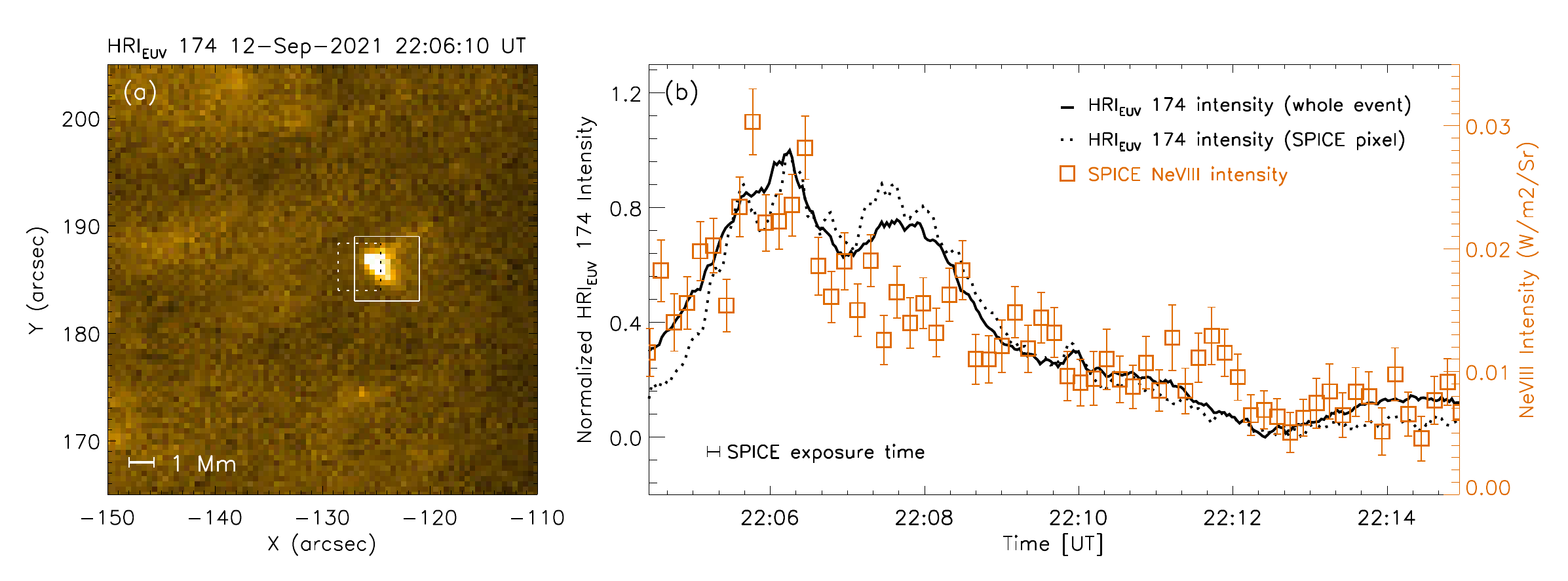}
\caption{EUI brightening (E-2). Same as Fig.~\ref{FigE12}, but for E-2 on 12 September 2021, this time showing the Ne~{\sc viii} intensity evolution (orange) in panel (b). SPICE exposure duration is 10~s in this case. An animation of panel (a) is available.}
\label{FigE22}
\end{figure*}

\section{Results}
\label{sec-res}
\subsection{23 February 2021: Event E-1}

One EUI brightening is recognized in this data set. In Fig.~\ref{FigE11} we show the time-slice images at the second spatial position, where the EUI brightening is most obvious. There, a brightening appeared for about 40~s in the upper part of the slit. Panels (a) and (b) of Fig.~\ref{FigE11} show the time-slice plots of \hrieuv\ data at full and reduced (to SPICE) resolution, respectively. The degradation in panel (b) is applied in both time and space directions by selecting \hrieuv\ data close in time to each SPICE spectrum and averaging pixels to resize them to the SPICE-binned spatial scale ($4\text{\arcsec} \times 4.392$\arcsec). The appearance and disappearance of the brightening, whose spatial location is indicated by the orange boxes, can be observed in both of these time-slice plots. This means that this EUI brightening should be detectable at the resolution of the SPICE data. 

Panel (a) of Fig.~\ref{FigE12} shows an \hrieuv\ image at about the peak time of EUI brightening E-1, giving an overview of its structure. The solid black line in panel (b) shows the corresponding light curve of this event (computed over the solid box shown in panel (a)) during the whole observing time. The \hrieuv\ 174 intensity (dashed line) is also calculated over the SPICE pixel (the dashed box shown in panel (a)), which shows no remarkable difference from the \hrieuv\ 174 intensity of the whole event. This EUI brightening is found to be very small (smaller than 4~Mm at maximum extent), but the combined heliocentric distance and high spatial resolution of \hrieuv\ make it possible to observe it in detail.

The time-slice plots of all spectral lines acquired by SPICE are also shown in Fig.~\ref{FigE11} together with the light curves showing the temporal change in the line intensities. Four pixels are binned along the slit direction in SPICE data. The error bars of these intensities, overplotted in orange, were calculated from the uncertainty of the Gaussian fitting parameters. In the \hrieuv\ light curve, the intensity of this EUI brightening increases by about $16\%$. However, in the spectral data from SPICE, only the O~{\sc vi} spectra show an increase around the time when the EUI brightening appears (from about 17:17 UT to 17:18 UT), although it is not as strong as the brightening in the \hrieuv\ light curve. When we consider that no obvious clue of the EUI brightening is found in Ne~{\sc viii}, this might indicate that this EUI brightening has a sufficient emission measure (to be detected by SPICE) only at temperatures close to the typical temperature of O~{\sc vi} (0.3~MK) and none, or very little, at coronal temperatures. The light curve of O~{\sc vi} in Fig.~\ref{FigE12} shows that SPICE may have missed the time at which the \hrieuv\ intensity reached its peak due to the time cadence of 20~s, which is almost half of the lifetime of this EUI brightening. 
In Fig.~\ref{FigE13} we show the spectra of the O~{\sc vi} line for this case at four different times, demonstrating the increased emission around the appearance time of the brightening. The radiance of the O~{\sc vi} line peaks at time 3 (see the legend in Fig.~\ref{FigE13}).

\begin{figure}[tbp]
\centering 
\includegraphics[width=1.\hsize]{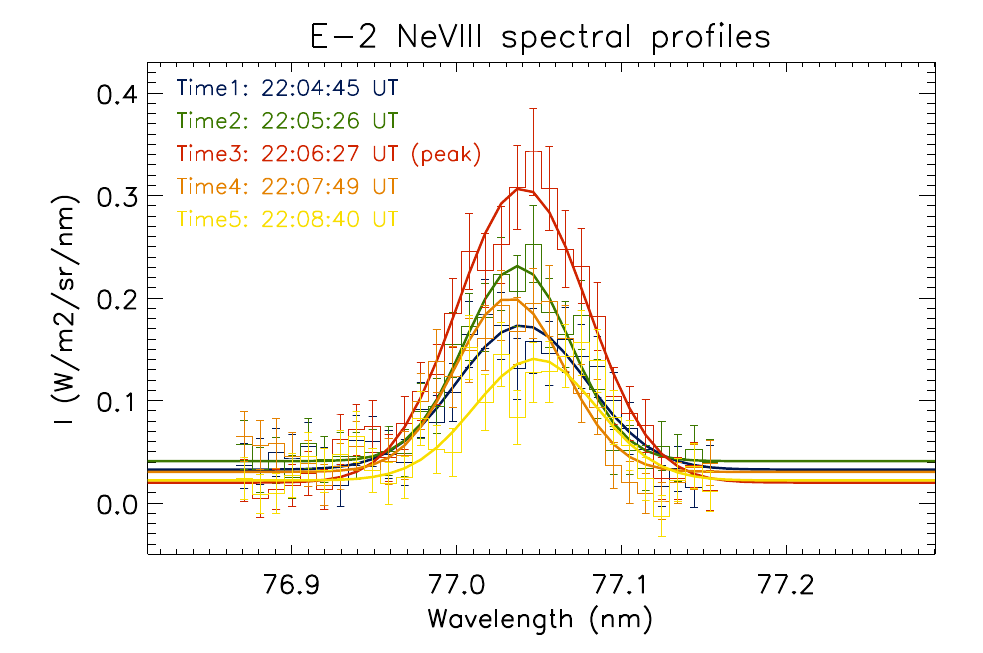}
\caption{E-2. Ne~{\sc viii} spectral profiles (histograms with error bars) and the Gaussian fitting results (thick lines) at five different times during the lifetime of E-2. }
\label{FigE23}
\end{figure}

\subsection{12 September 2021: Event E-2}

In the time-slice images showing the three time periods covered by \hrieuv\ data (see Table~\ref{tab1}) in data set 2, two EUI brightenings occur at the position of the SPICE slit. Event E-2 appeared in the time series from 22:04 to 22:10~UT on 12 September, and event E-3 appeared in the series from 00:19 to 00:23~UT on 13 September. Compared with E-1, the two EUI brightenings in this data set lasted longer (i.e., 6~min for E-2 and 4~min for E-3, instead of less than a minute for E-1) and were observed by SPICE at higher time cadence (i.e., 10~s instead of 20~s) and with a longer exposure time (i.e., 10~s instead of 4.7~s)

As in the case of E-1, in panels (a) and (b) of Fig.~\ref{FigE21}  we show \hrieuv\ time-slice plots with both original ($\sim0.5$\arcsec\,in space and $\sim2$~s in time) and degraded ($\sim4$\arcsec\,in space and $\sim10$~s in time) image scales to verify that the target EUI brightening can still be distinguished at the SPICE resolution. The bright trace left by this EUI brightening (in the orange boxes) and the corresponding enhancement in the light curve are still visible. Fig.~\ref{FigE22} shows the \hrieuv\ 174 image of this structure at its peak time (panel (a)) along with the light curves (panel (b)) of the small regions in the solid box (covering the whole event) and the dashed box (covering the binned SPICE pixel) in panel (a). This brightening peaked at about 22:06~UT. We also show the integrated line radiances of Ne~{\sc viii} at the position of the EUI brightening (orange squares in panel (b)). The light curves are roughly synchronous. 

Fig.~\ref{FigE21} also shows the time-slice plots of the integrated line radiances of all lines provided by SPICE and their corresponding light curves. This EUI brightening cannot be detected in all lines. In Ne~{\sc viii} (panel (i)), the light curve peaks at the same time as the \hrieuv\ light curve (22:06~UT), while the light curve of O~{\sc vi} (panel (h)) seems to have a broader peak that is also roughly centered around 22:06~UT.
On the other hand, the light curves of N~{\sc iv} (panel (e)), S~{\sc v} (panel (g)), and O~{\sc iv} (panel (f)) might peak at an earlier time. Unfortunately, SPICE observation only started at the beginning of the plotted interval, which probably fails to fully cover the rise phase of this event. 
The change with time in radiance of the Ne~{\sc viii} spectra is also shown in Fig.~\ref{FigE23}. The peak time difference between the different spectral lines may indicate that the temperature of this EUI brightening changes with time. Here, cooler lines such as C~{\sc III} and Ly$\beta$ show no obvious variation during the whole observation time. 

\begin{figure*}[tbp]
\centering 
\includegraphics[width=1.0\textwidth]{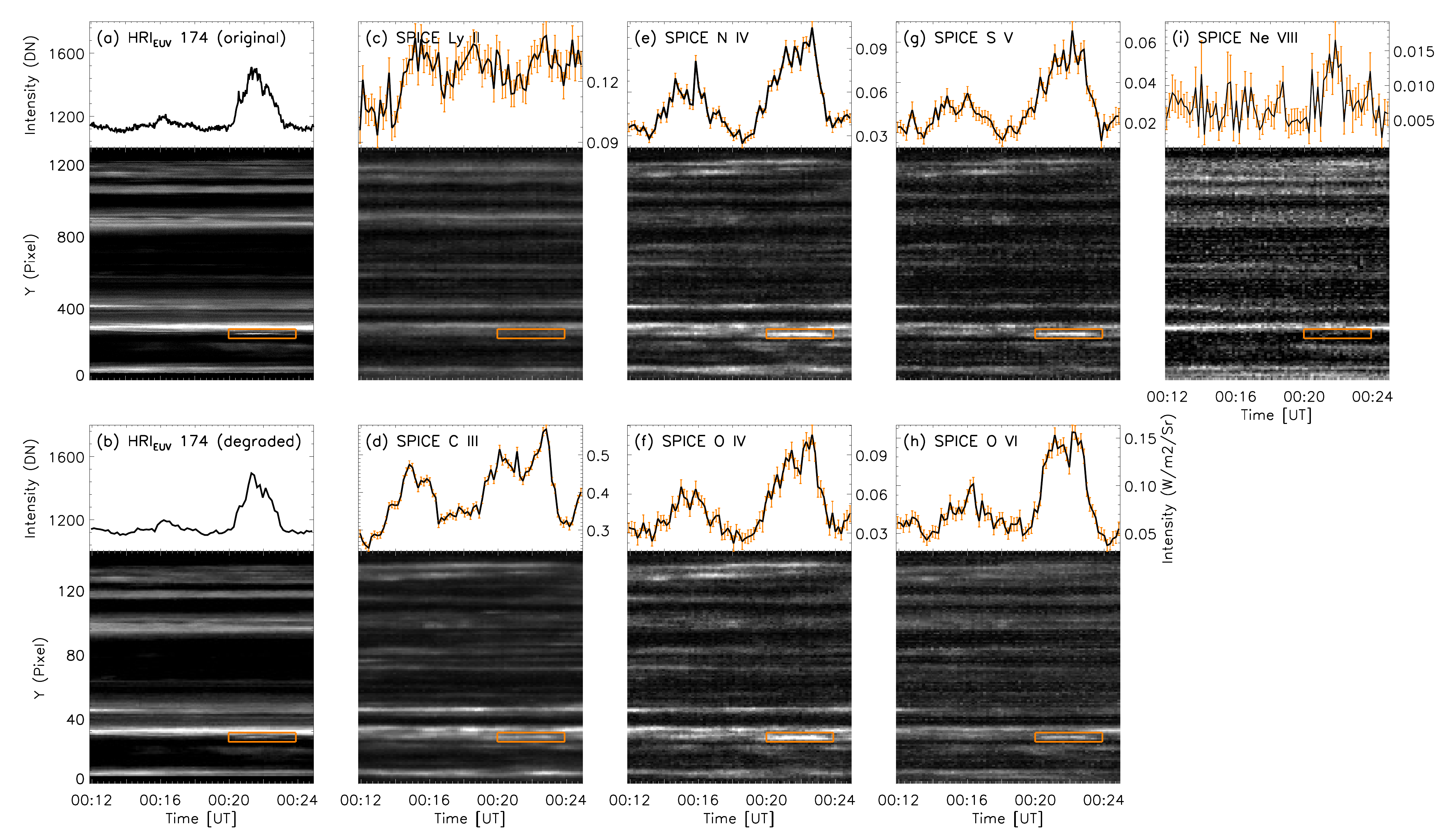}
\caption{Same as Fig.~\ref{FigE11}, but for E-3 on 13 September 2021. It shows time-slice plots and light curves from the \hrieuv\ data at full and reduced (to SPICE) resolution and from Gaussian fitting results of the Ly$\beta$, C~{\sc iii}, N~{\sc iv}, O~{\sc iv}, S~{\sc v}, O~{\sc vi,} and Ne~{\sc viii} spectral lines.}
\label{FigE31}
\end{figure*}

\begin{figure*}[tbp]
\centering 
\includegraphics[width=1.0\textwidth]{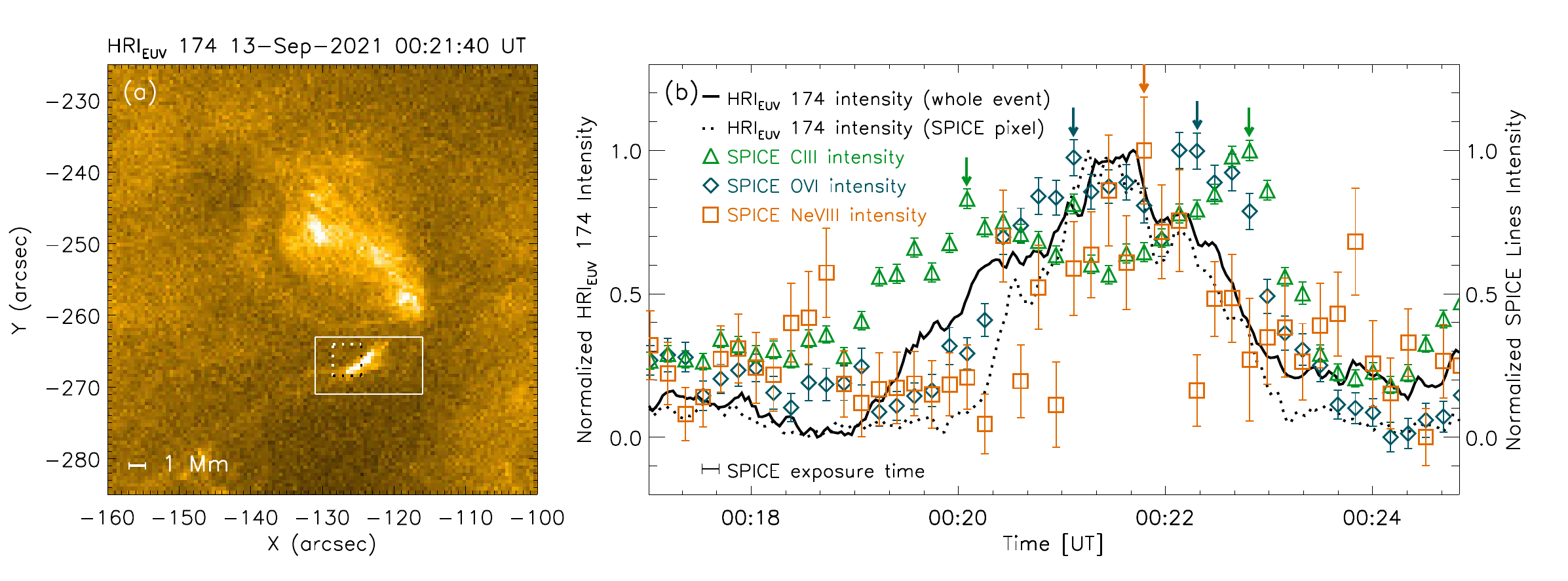} 
\caption{Same as Fig.~\ref{FigE12}, but for E-3 on 13 September 2021 and this time showing the C~{\sc iii} (green), O~{\sc vi} (blue) and Ne~{\sc viii} (orange) intensity evolution (normalized). Their peaks are marked by arrows in different colors. The SPICE exposure duration is 10~s in this case. An animation of panel (a) is available.}

\label{FigE32}
\end{figure*}

\begin{figure*}[tbp]
\centering
\includegraphics[width=1.0\textwidth]{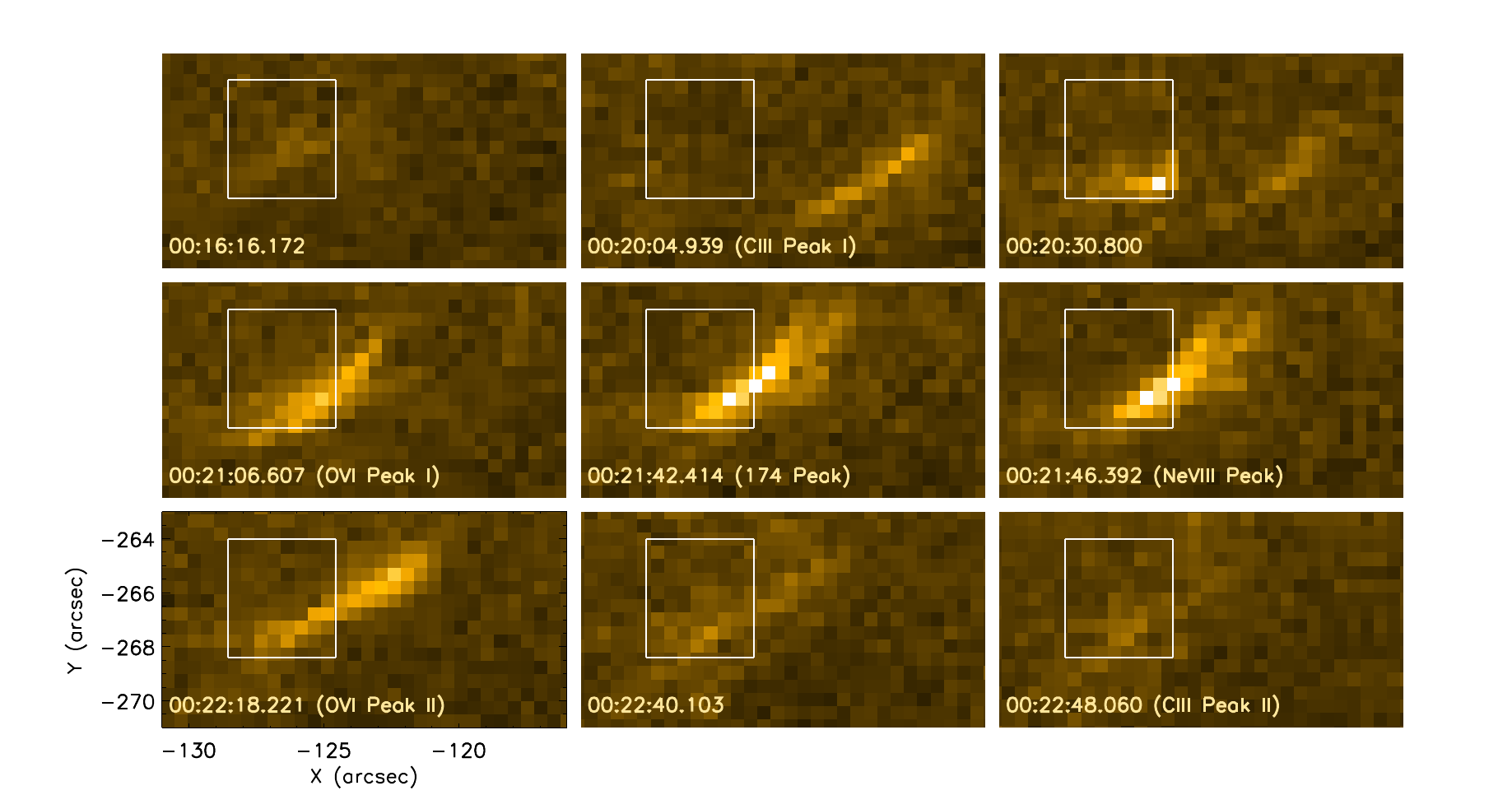} 
\caption{Event E-3 at nine different times, including the peak times of the intensities of \hrieuv\ 174, SPICE Ne~{\sc viii}, C~{\sc iii} and O~{\sc vi}  . The white boxes show the binned SPICE pixel covering the EUI brightening.}
\label{FigE34}
\end{figure*}

\subsection{13 September 2021: Event E-3}

SPICE also provided data at high time-cadence for E-3. 
Fig.~\ref{FigE31} shows the time-slice images of \hrieuv\ at full resolution (panel (a)) and at a resolution that was reduced to the spatial and temporal resolution of SPICE (panel (b)). The degraded time-slice image and corresponding light curve show that this EUI brightening can be detected at the spatial and temporal resolution of SPICE. 

This EUI brightening is a loop-like event south of a larger brightening. In panel (a) of Fig.~\ref{FigE32} , we show the image of this EUI brightening and its neighborhood at the time when it reached its peak intensity in the \hrieuv\ 174 channel. We selected the region covering the EUI brightening, as shown by the white box in Fig.~\ref{FigE32} panel (a), to calculate the \hrieuv\ light curve in panel (b) of Fig.~\ref{FigE32} . It shows the intensity evolution of the whole event. In addition, the light curves from the SPICE lines and from \hrieuv\ (dashed line) were calculated in the area marked by the dashed box and are also plotted in panel (b). The \hrieuv\ light curves of the whole event and of the area within the SPICE-binned pixel differ significantly only around 00:20~UT because a small brightening occurs before the main event, as shown in the top middle panel of Fig.~\ref{FigE34} (see the discussion below).

The evolution of event E-3 is shown in Fig.~\ref{FigE34}, where the regions we used to calculate the light curve in panel (b) of Fig.~\ref{FigE32}  are plotted for nine different times, including the times when all the spectral lines reached their peak intensities. These figures show that this brightening first appeared as a bright confined loop and then extended northwest, forming an elongated structure. Before its occurrence, another fainter loop can be seen in upper panels, west of the SPICE slit, moving from northwest to southeast. This may indicate that this bright loop transports energy to one footpoint and triggers the EUI brightening that is captured in the SPICE slit. The propagation of brightenings was also reported in \citet{2021A&A...656L..16M}. % Mandal et al 2021
A similar evolution can be found in jet observations \citep{2021A&A...656L..13C}, % Chitta et al. 2021b
where the reconnection occurs between a local bipolar field and an open field or a field connecting to a far footpoint \citep{1977ApJ...216..123H, 1992PASJ...44L.173S}. % Heyvaerts et al. 1977, Shibata et al. 1992 
Some small-scale jets are found to be related with footpoint brightenings. They reach temperatures of $10^5$~K at least \citep{2014Sci...346A.315T}.

\begin{figure}[tbp]
\centering 
\includegraphics[width=1.\hsize]{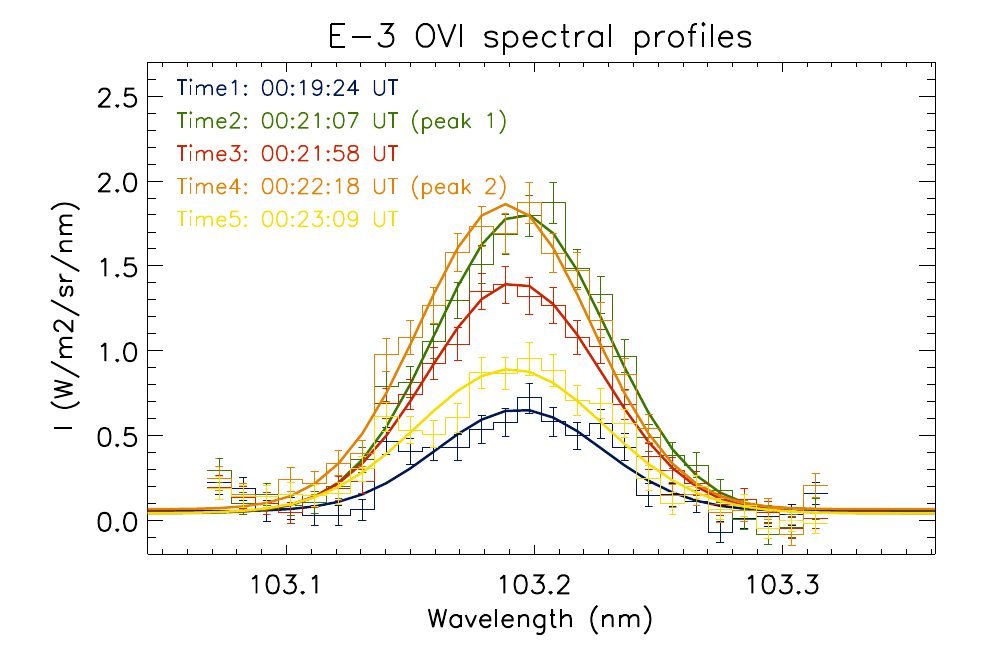}
\caption{E-3. O~{\sc vi} spectral profiles (histograms with error bars) and the Gaussian fitting results (thick lines) at five different times during the lifetime of E-3.}
\label{FigE33}
\end{figure}

Gaussian fitting results of the SPICE spectral lines are shown in panels (c -- i) in Fig.~\ref{FigE31}. They reveal that at about the peak time of the event, the Ne~{\sc viii} and some TR lines have enhanced emission during this EUI brightening. In the light curves showing the temporal change of the line intensities, the EUI brightening emission in some TR lines shows two peaks around the peak time of the \hrieuv\ data and of Ne~{\sc viii}  (00:22~UT) (panels (a) and (i) in Fig.~\ref{FigE31}). This is shown in greater detail in Fig.~\ref{FigE32} and   is particularly evident in O~{\sc vi} in Fig.~\ref{FigE33}. That the EUI brightening has different peak times in different spectral lines indicates an evolution of the plasma temperature during the event. The O~{\sc vi} and C~{\sc iii} lines both show two peaks that are more distant in time for C~{\sc iii} and closer for O~{\sc vi} to the time when Ne~{\sc viii} reaches its only peak (see Fig. \ref{FigE32}). This is an indication that the temperature of this EUI brightening increases from the typical emission temperature of C~{\sc iii} ($\log T\rm{[/K]}=4.8$) to that of O~{\sc vi} ($\log T\rm{[/K]}=5.5$) and then continues to rise to Ne~{\sc viii} ($\log T\rm{[/K]}=5.8$), which is the temperature of the lower corona. After the Ne~{\sc viii} peak time, the EUI brightening started to cool down to TR temperatures, which explains the appearance of the second emission peak in O~{\sc vi} and C~{\sc iii} lines. Before the occurrence of this EUI brightening, at about 00:16~UT, another enhancement is clearly visible in TR lines (see Fig.~\ref{FigE31}). This might be another related small brightening or a precursor of this EUI brightening. Although the \hrieuv\ 174 light curves also show this structure, the intensity is not comparable with the EUI brightening we study here. Similarly, the upper left panel in Fig.~\ref{FigE34} shows that a faint brightening was covered in the SPICE pixel, minutes before the EUI brightening appeared.
This first brightening may only have reached lower temperatures and may even suggest a process of energy accumulation prior to the EUI brightening.

\section{Discussion and conclusion}
\label{sec-con}

We have analyzed three EUI brightenings that were observed simultaneously by \hrieuv\ and SPICE on board Solar Orbiter. These three events were observed at two widely different times, and different types of data were provided by SPICE. The first event occurred on 23 February 2021, and we used a sequence of three-step SPICE rasters with a time cadence of 20~s. This event lasted for approximately 40~s. The other two events occurred on 12 and 13 September 2021, during which SPICE observed continuously in sit-and-stare mode with a cadence of 10~s.
Both events lasted for several minutes. Here we point out that the scales of all three EUI brightening events are close to that of the smallest structure SPICE can detect. Their identification therefore depends on \hrieuv\ high-resolution imaging.

Combining the \hrieuv\ imaging observations with the spectral data provided by SPICE,
%, using Gaussian fit and estimating errors based on reliable parameters, 
we find that the first event can only be observed in O~{\sc vi}, while the other two longer-duration EUI brightenings are detectable not only in TR lines, but also in Ne~{\sc viii}. In both cases, the Ne~{\sc viii} emission shows a single peak that coincides with the peak seen in \hrieuv. This may imply that some EUI brightenings barely reach coronal temperatures. This scenario is in one case supported by the fact that we observe two peaks in the light curve of TR lines, while the light curve of Ne~{\sc viii} only reaches one peak in between the double peaks from the lines formed at lower temperatures. 
This can help us to infer the process by which the EUI brightening heats to the temperature of lower corona and then decreases. Compared with the DEM results in \citet{2021A&A...656L...4B}, % Berghmans et al. 2021 
who reported that EUI brightenings can reach coronal temperature, our cases are more likely to have a lower temperature. A similar result can be found in \citet{Dolliou_tbs}, % Dolliou et al. submitted to this issue, published
who conducted a time-lag analysis of EUI brightenings. This measures the time delays between the light curves of different AIA channels. The EUI brightenings they studied are more likely to be dominated by cool plasma that does not reach coronal temperature. In addition to thermal evolution, the different peak times might also be caused by spatial movements of the EUI brightenings. When we consider that the SPICE slit is relatively thin and the EUI brightenings are active, it might be possible for brightenings to partially move in and out of the SPICE slit.

To explore whether our results here can be generalized to more EUI brightenings, more samples need to be studied. This study clearly shows that SPICE does have the capability of observing EUI brightenings, although it requires imaging observation with higher spatial resolution such as those from EUI to clearly identify these small brightenings. 
Because EUI brightenings evolve so fast, observations with a higher time cadence (e.g., sit-and-stare mode) would be helpful. Although the SPICE slit in sit-and-stare mode is fixed at one single position, which misses most small brightenings, there is a good chance for us to observe at least one of them along the slit, considering the high frequency of EUI brightening occurrences. With Solar Orbiter now in its nominal science phase, it is expected that more events will be captured by both \hrieuv\ and SPICE. Many of these observations will also benefit from higher spatial resolution if they are carried out close to the science perihelia. In addition, other spectrometers also provide very valuable data \citep[e.g., IRIS,][]{2014SoPh..289.2733D}, which can be combined with the high-resolution \hrieuv\ data to identify EUI brightenings. EUI brightenings detected in a 29.4~min EUI data set taken on 8 March 2022 are also studied with coordinated IRIS observations to understand their response in TR lines by \citet{Nelson_tbs_IRIS}. % Nelson et al. iris submitted to this issue
Some but not all of the EUI brightenings covered in that study show a clear cotemporal and cospatial signature in the Mg~{\sc ii} and Si~{\sc iv} lines. No typical response of EUI brightenings in the TR was concluded from that study. By studying more small-scale brightening events, we might proceed in understanding their heating mechanism and their contribution to coronal heating. 

\begin{acknowledgements}

\color{blue}
The authors thank the referee for helpful comments that have improved this paper.
Solar Orbiter is a space mission of international collaboration between ESA and NASA, operated by ESA.
The development of SPICE has been funded by ESA member states and ESA. It was built and is operated by a multi-national consortium of research institutes supported by their respective funding agencies: STFC RAL (UKSA, hardware lead), IAS (CNES, operations lead), GSFC (NASA), MPS (DLR), PMOD/WRC (Swiss Space Office), SwRI (NASA), UiO (Norwegian Space Agency).  
The EUI instrument was built by CSL, IAS, MPS, MSSL/UCL, PMOD/WRC, ROB, LCF/IO with funding from the Belgian Federal Science Policy Office (BELSPO/PRODEX PEA C4000134088); the Centre National d’Etudes Spatiales (CNES); the UK Space Agency (UKSA); the Bundesministerium f\"{u}r Wirtschaft und Energie (BMWi) through the Deutsches Zentrum f\"{u}r Luft- und Raumfahrt (DLR); and the Swiss Space Office (SSO).
L.P.C. gratefully acknowledges funding by the European Union (ERC, ORIGIN, 101039844). Views and opinions expressed are however those of the author(s) only and do not necessarily reflect those of the European Union or the European Research Council. Neither the European Union nor the granting authority can be held responsible for them.
S.P. acknowledges the funding by CNES through the MEDOC data and operations center. 

\color{black}
\end{acknowledgements}


\bibliographystyle{aa} % style aa.bst

%\bibliography{ZH-bibtex} % your references Yourfile.bib

\begin{appendix}

\section{Data alignment}
\label{data-align}

\begin{figure*}[tbp]
\centering 
\includegraphics[width=1.\hsize]{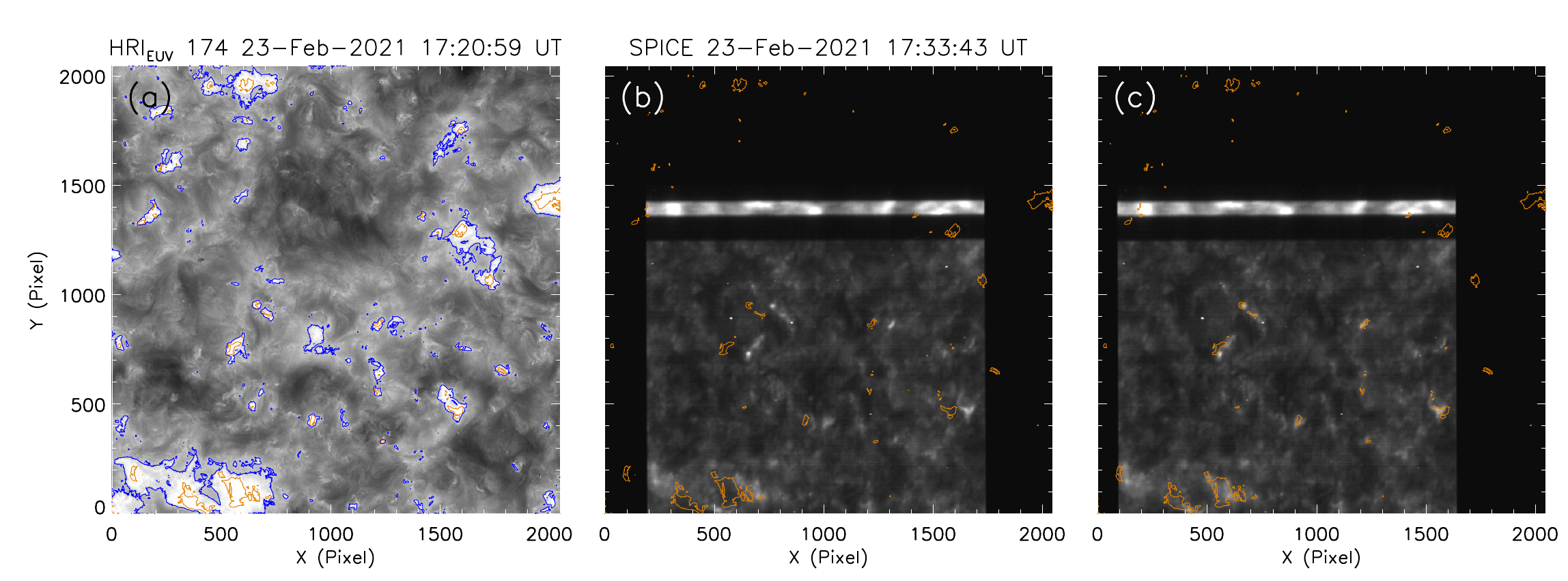} 
\caption{First step of data coalignment. Panel (a): \hrieuv\ image taken on 23 February 2021 at 17:20:59~UT with intensity isocontours of 1500 (blue) and 2000 (orange) DN/s. Panel (b): Corresponding (reprojecting the SPICE data over the \hrieuv\ field of view by using the WCS keywords in the file headers) intensity map of Ne~{\sc viii} line radiance from the SPICE context raster starting at 17:33:43~UT with the same \hrieuv\ contours of 2000 DN/s as in panel (a). Panel (c): Same as panel (b), but the intensity map is manually coarse-aligned to the \hrieuv\ intensity.}
\label{FigA1}
\end{figure*}

\begin{figure*}[tbp]
\centering 
\includegraphics[width=1.\hsize]{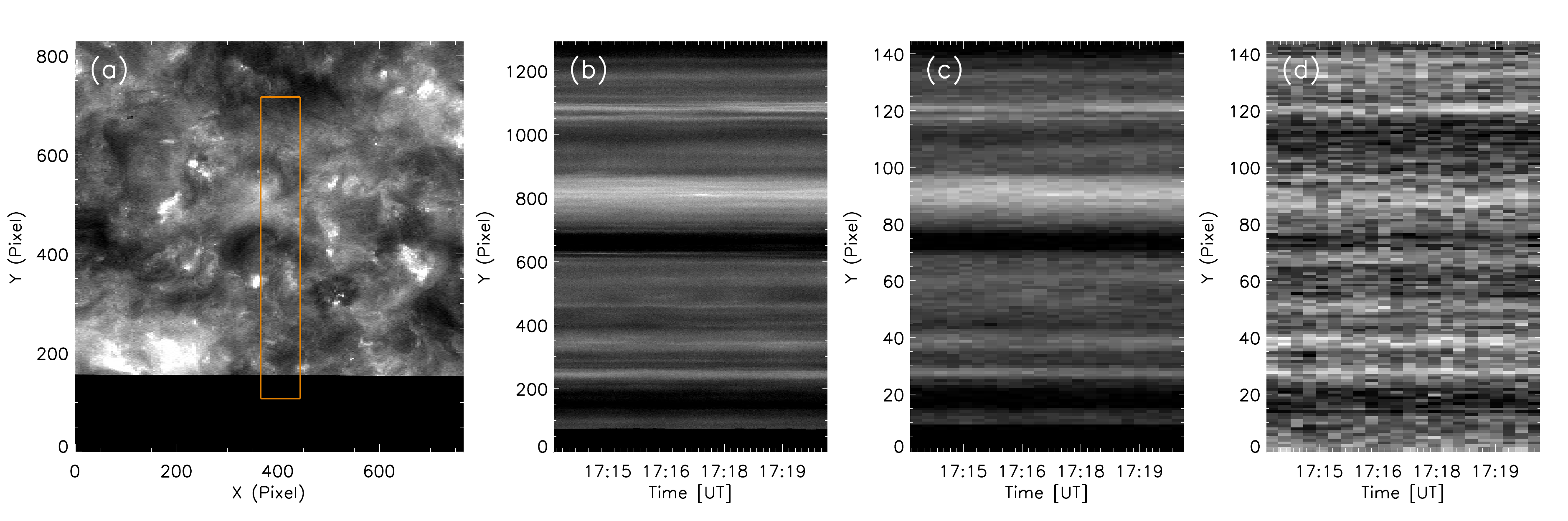} 
\caption{Example of the second step of data coalignment. Panel (a): \hrieuv\ image reprojected over the SPICE context raster using the WCS keywords in the file header. The orange box shows the region in which the SPICE slit was thought to be based on the results from the first coarse coalignment step. Panel (b): \hrieuv\ time-slice image with full resolution by selecting one slit position shown in panel (a). Panel (c): Same as panel (b), but the resolution is degraded to the SPICE imaging resolution. Panel (d): Time-slice image of the Ne~{\sc viii} amplitude from Gaussian fitting results.}
\label{FigA2}
\end{figure*}

The header information in the data files only provides approximate information about the alignment of \hrieuv\ and SPICE, mostly because the two instruments are mounted on different panels of the spacecraft and thermal variations due to the highly elliptic orbit affect them differently.
In addition to this, there is also the complication, intrinsic to a scanning spectrometer, that different from a snapshot from an imager, an image built by scanning the slit for a long time (more than one hour) is strongly affected by solar rotation. Although the last effect can easily be computed and accounted for, we would still have the problem that the context rasters (those whose field of view is large enough for a meaningful correlation analysis) are taken at different times than the sit-and-stare or fast three-step rasters used in our analysis.
For these reasons, we used a two-step approach that allows the direct alignment of the sit-and-stare (or high-cadence three-step rasters) SPICE data with the \hrieuv\ images.

The first step consists of visually refining the coalignment between the context raster (reprojected over the \hrieuv\ data, see below) and the \hrieuv\ data. This was done ignoring solar rotation, as the scope is to get close enough for the next alignment step.
Figure~\ref{FigA1} shows the last \hrieuv\ image of the 23 February data set (taken at 17:20:59~UT) with the intensity isocontours of 1500 and 2000 DN/s overplotted, showing the outlines of the brightenings (panel (a)). Panel (b) shows the image from the SPICE context raster scan starting at 17:33:43~UT reprojected over the \hrieuv\ image using the world coordinate system (WCS\footnote{\url{https://fits.gsfc.nasa.gov/fits_wcs.html}} ) keywords in the L2 FITS header.
The same isocontours of 2000 DN/s as in panel (a) are also plotted in this projected Ne~{\sc viii} context raster. Panel (c) shows the same image as in panel (b) after the visual refinement is applied to the coalignment (first coarse coalignment step). 
The residual mismatch due to solar rotation is well visible as a compression of the image in the scanning direction because the field of view rotates in the same direction as the slit scanning. 

\begin{figure*}[tbp]
\centering 
\includegraphics[width=1.\hsize]{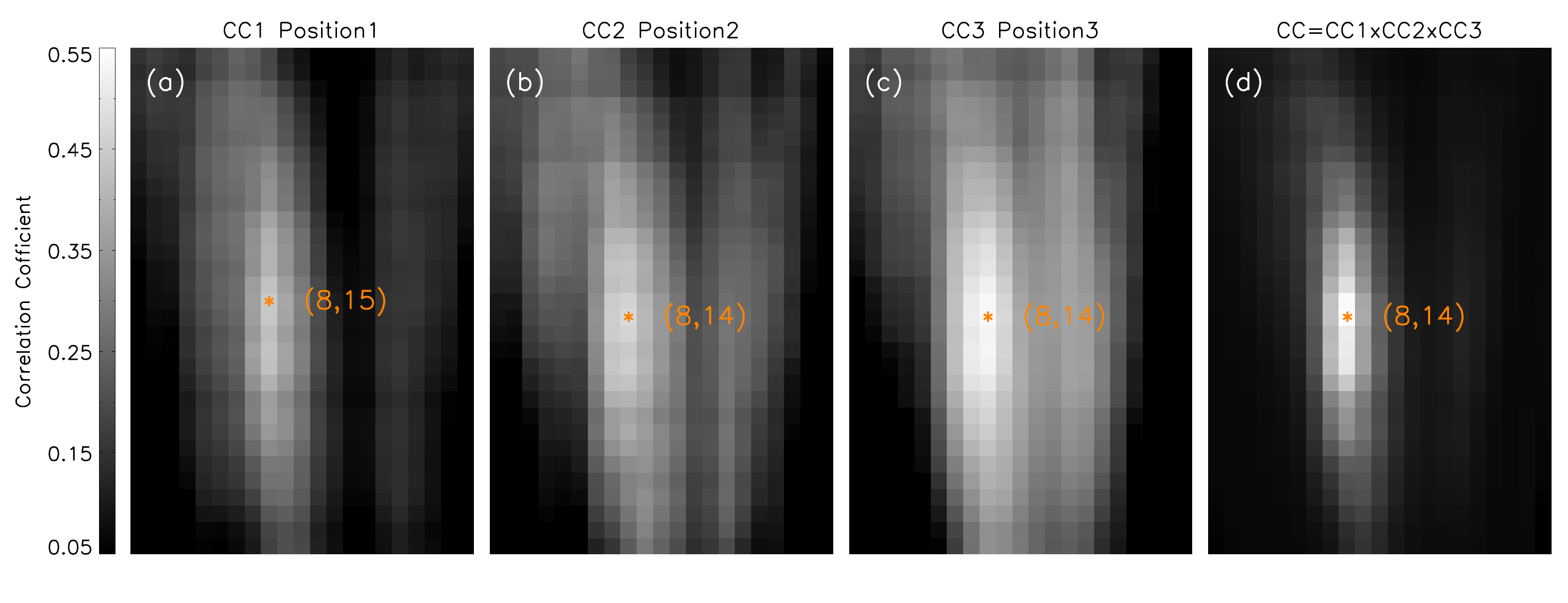}
\caption{Correlation coefficient maps. Panels (a)-(c): Correlation coefficient maps between time-slice plots of the degraded \hrieuv\ time-slice image and the Ne~{\sc viii} amplitude time-slice image at three positions. The color bar shows the range of correlation coefficients plotted on the left side. Panel (d): Map of the production of the correlation coefficient at three positions. The orange stars mark the slit position with the highest correlation coefficient. Its pixel coordinates are denoted in parentheses.}
\label{FigA3}
\end{figure*}

The second alignment step makes use of the time-series data (three time series, one for each raster location for the 23 February 2021 data). 
In this step, we aligned \hrieuv\ and SPICE data by comparing time-slice images. In the time covered by the \hrieuv\ and SPICE high-cadence data, we fit the spectra of Ne~{\sc viii}, which is closest in temperature sensitivity to the \hrieuv\ bandpass over the average corona, with a single-Gaussian function and made a time-slice plot of the obtained line amplitude. 
To make the time-slice image of \hrieuv\ data cotemporal and cospatial to the SPICE data, we first selected the \hrieuv\ images closer in time to the individual SPICE spectra and then downsized them to the SPICE spatial scale ($4\text{\arcsec} \times 4.392$\arcsec\,after binning along the slit) by averaging pixels together. In Fig.~\ref{FigA2} we show an example from the data set containing event E-1. Panel (a) shows an \hrieuv\ image reprojected over the SPICE context raster. Panels (b) and (c) show time-slice images of full- and degraded-resolution \hrieuv\ data, respectively. The spatial location of the slice at this stage is based on the result from the first step of the alignment process. Finally, panel (d) shows the time-slice image from the Ne~{\sc viii} line amplitude. 
The orange box in panel (a) shows the region over which we explored the correlation of all possible time-slice images from \hrieuv\ with the SPICE time-slice image. 
For each of the possible time-slice images, we calculated the correlation coefficient and then selected the location of maximum correlation. In the case of the 23 February data set, this was done for each of the three raster positions. Panels (a), (b), and (c) in Fig.~\ref{FigA3} show the maps of the correlation coefficients obtained for the three raster positions while moving the alignment position over $21\times31$ \hrieuv\ full-resolution pixels. Panel (d) shows the map of the product of the three individual correlation coefficients. The position of the peak provides the final adjustment to the alignment found with the first alignment step. In the case of the data sets of September 2021, the analysis was carried out for the three individual data sets corresponding to each of the three raster scan positions.

\section{Error analysis}
\label{error}

\begin{figure}[tbp]
\centering 
\includegraphics[width=1.\hsize]{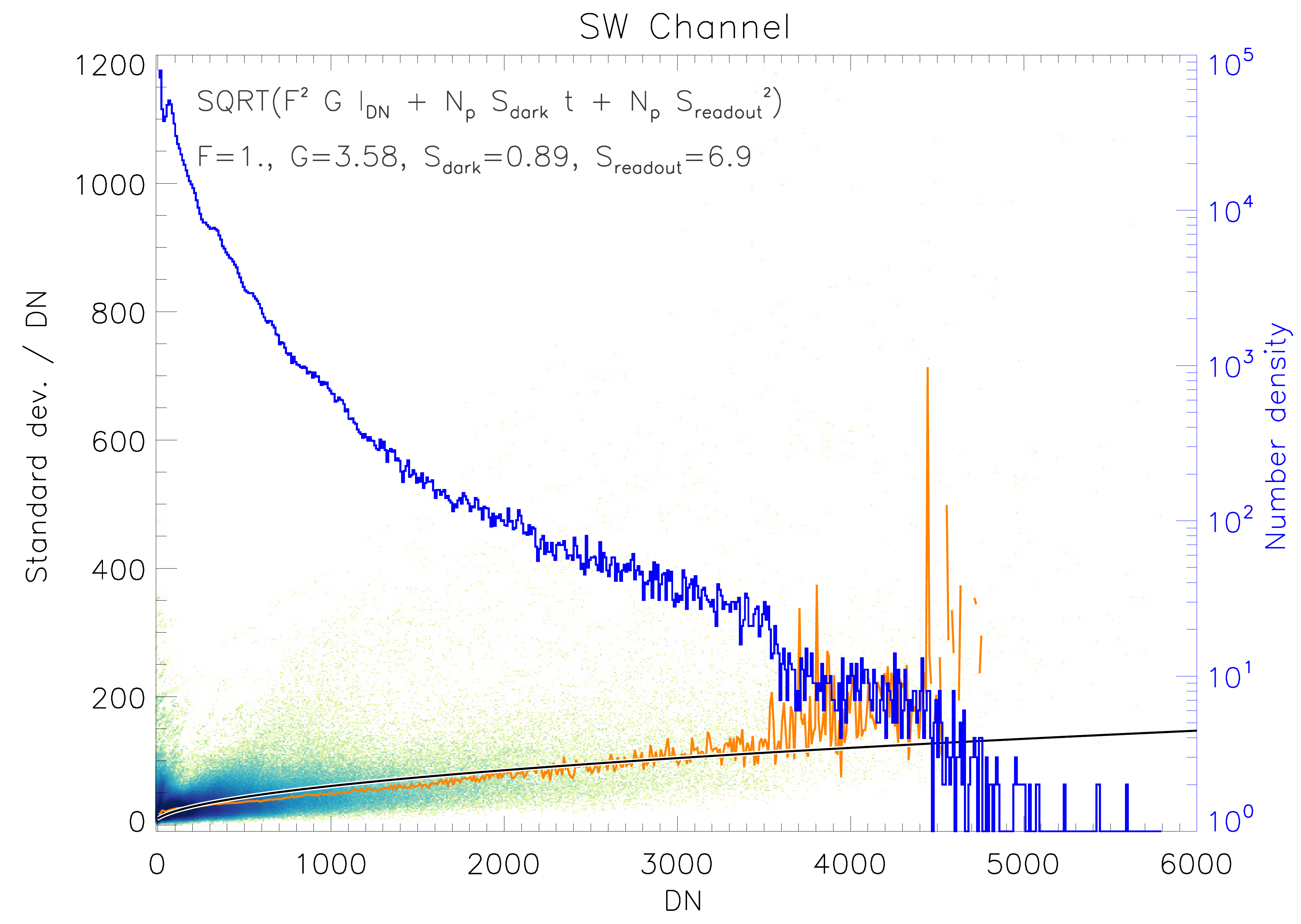} 
\caption{Noise estimate for unbinned SW data. The 2D histogram shows the standard deviation as a function of signal across the image. The orange line is the median to the data point in ten DN bins. The black line is the outcome of Eq.~\eqref{eq:sigmadn} for $F$=1. The blue histogram shows the data number density.}
\label{FigB1}
\end{figure}

\begin{figure}[tbp]
\centering 
\includegraphics[width=1.\hsize]{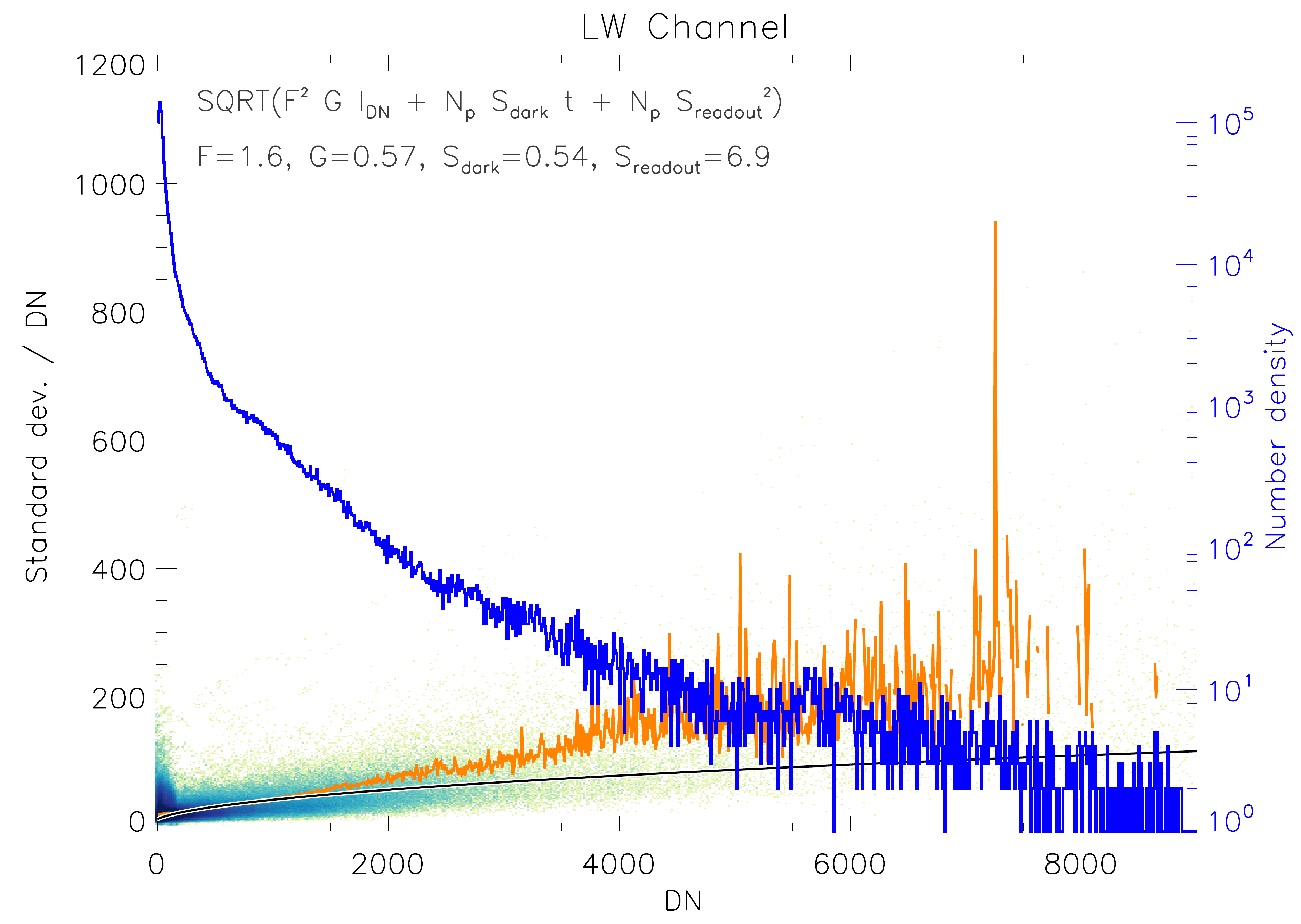} 
\caption{As in Fig.~\ref{FigB1}, but for the LW channel. The data are fit by $F$=1.6.}
\label{FigB2}
\end{figure}

For an average signal $\overline{I}_{DN}$ (average accumulated DN per detector pixel over an exposure time of $t$ seconds), the uncertainty due to photon noise over a number $n_p$ of pixels is 
$\sigma_{ph}=\sqrt{F^2 \overline{I}_{DN}\,n_p / G}$, where $G$ is the camera conversion gain in DN/photons, and $F$ is the intensifier noise factor, accounting for the amplification effect of the intensifier. It usually has a value between 1.3 and 2 \citep{2003SPIE.4796..164D}. % Denvir & Conroy, 2003) 
Here, it also covers some of the uncertainties in the determination of $G$. 

Then the uncertainty (in DN) due to photon noise of the signal $\overline{I}_{DN}\,n_p$ is $\sigma_{DN}=\sqrt{F^2\,\overline{I}_{DN}\,n_p\,G}$. 
The total uncertainty is then obtained by square-summing the photon noise component to the readout noise $\sigma_{read}$ (square root summed when binning over more pixels) and to the dark noise (which is equal to the square root of the dark current $n_p I_{dark}$, in units of DN/s). 

Thus, 

\begin{equation}
  \Sigma_{DN}=\sqrt{F^2\overline{I}_{DN}\,n_p\,G +n_p\,\sigma^2_{read}+n_p\,I_{dark}\,t}.  
\end{equation}

Estimates of $G$, $\sigma_{read}$ , and $I_{dark}$ have been obtained by the SPICE team through on-ground and in-flight calibration activities and are listed in Table \ref{tabA1}.
Since SPICE on-board binning returns the signal in DN summed over the number of binned pixels, it is more practical to consider the uncertainties over this signal $I_{DN}=\overline{I}_{DN}\,{n_p}$ , which can be written as

\begin{equation}
  \label{eq:sigmadn}
  \Sigma_{DN}=\sqrt{F^2 I_{DN}\, G +n_p\,\sigma^2_{read}+n_p\,I_{dark}\,t}. 
\end{equation}

To verify this approach and to determinate the value of $F$, we took an uncalibrated (L1) full detector observation acquired on 21$^{\rm }$ December 2021 at 15:08:34~UT using the wide 30\arcsec\,slot and corrected it by subtracting a matching dark from 
15$^{\rm }$ December 2021 at 08:32:20~UT and by dividing it by the flat-field matrix. These data belong to SPICE data release 2.0\footnote{\url{https://doi.org/10.48326/idoc.medoc.spice.2.0}}. The data were also corrected for cosmic spikes and hot pixels. The procedure replaced about 2000 (1000) pixels in the SW (LW) image. This number is negligible with respect to the total number of pixels in the ($1024\times1024$) data array. As the SPICE spatial resolution is about 6 to 7 pixels and the slot adds additional blur in the spectral direction as well, we consider that the noise level associated with the average signal over a $3\times3$ pixel area in such an image can be estimated as the standard deviation over the same $3\times3$ pixel area. 

In order to prevent the intensity gradient from altering the results, a smoothed (two passes of a $7\times7$ box car) copy of the image was subtracted from the image before we estimated the standard deviation. The signal value to which the standard deviation is associated was instead obtained from the original image. The procedure was applied to each pixel in the data arrays.

Figures~\ref{FigB1} (for the SW channel) and~\ref{FigB2} (LW channel) show a scatter plot of the standard deviation versus the signal. The thick orange line is the median passing through the points (within bins of ten DN). The black line represents the theoretical values from Eq.~\eqref{eq:sigmadn}. Finally, the blue histogram shows the number density of the data points over bins of 10 DN. A good match to the median values can be obtained with reasonable values of $F$ for the large majority of the intensity range.

\begin{table}[ht]
    \caption{Error estimate parameters}
    \label{tabA1}
    \centering
    \begin{tabular}{cccccc}
         \hline
         & $F$ & $G$ (DN/phot) & $\sigma_{read}$ (DN) & $I_{dark}$ (DN/s)\\
         \hline
         SW & 1 & 3.58 & 6.9 & 0.89\\
         LW & 1.6 & 0.57 & 6.9 & 0.54\\
         \hline
    \end{tabular}
\end{table}

The uncertainties can be calculated directly for calibrated L2 data by using the header keyword RADCAL. This contains the inverse average (over the spectral window under consideration) calibration factor $\alpha$ in units of $[\text{DN}/(\text{W}/\text{m}^2/\text{sr}/\text{nm})]$, which allows retrieving the total DN in the pixel (or in the original group of pixels in the case of data binned on board, as it is already multiplied by the onboard binning $n_p$) such that
$I_{DN}=\alpha \, I_{cal}$.

In fact,

\begin{equation}
\Sigma_{cal}=I_{cal}\frac{\Sigma_{DN}}{I_{DN}}=\frac{\Sigma_{DN}}{\alpha}=\frac{1}{\alpha}\sqrt{F^2 \alpha I_{cal}G +n_p\sigma^2_{read}+n_p I_{dark}t}
,\end{equation}

where the quantity $n_p$ is given by the keyword NBIN in the FITS header.
The above equation can be further generalized to cover the case when binning by a factor $n_u$ would be applied by averaging L2 calibrated data (whether binned on board or not). 
When the binning is made by the user, we have that 
$I_{DN}=\alpha \, I_{cal} n_u$. To calculate the uncertainties, we are interested in retrieving the total number of counts $I_{DN}$ within the binned area. 
Thus,

\begin{equation}
\Sigma_{cal}(\lambda)=\frac{1}{\alpha(\lambda)}\sqrt{\frac{F^2 \alpha(\lambda) I_{cal}G}{n_u} +\frac{n_p}{n_u}\sigma^2_{read}+\frac{n_p}{n_u} I_{dark}t}
.\end{equation}

Finally, the FITS headers as binary tables associated with the different windows also contain the full inverse calibration curve over wavelength (also called RADCAL).
The error calculations using the wavelength-dependent calibration factor $\alpha(\lambda)$ are relevant only in the case of large spectral windows and of full detector spectra, particularly in the case of the SW detector.

\end{appendix}
 
\end{document}